\documentclass[11pt,twoside]{article}
\usepackage{asp2014}

\aspSuppressVolSlug
\resetcounters

\begin{document}

\title{(O3-2) Data model as agile basis for evolving calibration software}

% the commas separating the authors should go after each author's name and before the corresponding affiliation number
\author{Hugo~Buddelmeijer,$^1$ Gijs~A.~Verdoes~Kleijn,$^1$ and Kieran~Leschinski$^2$}
\affil{$^1$OmegaCEN, Kapteyn Astronomical Institute, University of Groningen, Groningen, The Netherlands; \email{H.Buddelmeijer@rug.nl}}
\affil{$^2$Department of Astrophysics, University of Vienna,Vienna, Austria}
% remove/add authors as you need
\paperauthor{Hugo~Buddelmeijer}{hugo@buddelmeijer.nl}{0000-0001-8001-0089}{University of Groningen}{Kapteyn Astronomical Institute}{Groningen}{Groningen}{9700 AV}{Netherlands}
\paperauthor{Gijs~A.~Verdoes~Kleijn}{hugo@buddelmeijer.nl}{0000-0001-8001-0089}{University of Groningen}{Kapteyn Astronomical Institute}{Groningen}{Groningen}{9700 AV}{Netherlands}
\paperauthor{Kieran~Leschinski}{}{}{University of Vienna}{Department of Astrophysics}{Vienna}{}{}{Austria}
%\paperauthor{Sample~Author2}{Author2Email@email.edu}{ORCID_Or_Blank}{Author2 Institution}{Author2 Department}{City}{State/Province}{Postal Code}{Country}
% remove/add as you need

% leave these next few aindex lines commented for the editors to enable them. Use Aindex.py to generate them for yourself.
% first presenting author should be the first entry for bold-facing the author index page-reference
%\aindex{Buddelmeijer,~H.}
%\aindex{Verdoes~Kleijn,~G.~A.}
%\aindex{Leschinski,~K.}
% remove/add as you need

% leave the ssindex lines commented for the editors to enable them, use Index.py to suggest yours
%\ssindex{FOOBAR!conference!ADASS 2018}
%\ssindex{FOOBAR!organisations!ASP}

% leave the ooindex lines commented for the editors to enable them, use ascl.py to suggest yours
%\ooindex{FOOBAR, ascl:1101.010}

\begin{abstract}
We design the imaging data calibration and reduction software for MICADO, the First Light near-IR instrument on the Extremely Large Telescope.
In this process we have hit the limit of what can be achieved with a detailed software design that is primarily captured in pdf/word documents.

Trade-offs between hardware and calibration software are required to meet stringent science requirements. To support such trade-offs, more software needs to be developed in the early phases of the project: simulators, archives, prototype recipes and pipelines.
This requires continuous and efficient exchange of evolving designs between the software and hardware groups, which is hard to achieve with manually maintained documents.
This, and maintaining the consistency between the design documents and various software components is possible with a machine readable version of the design.

We construct a detailed design that is readable by both software and humans.
From this the design documentation, prototype pipelines and data archives are generated automatically.
We present the implementation of such an approach for the calibration software detailed design for the ELT MICADO imager which is based on expertise and lessons learned in earlier projects (e.g. OmegaCAM, MUSE, Euclid).

\end{abstract}
\section{Context: evolving hardware and calibration software to achieve cutting edge science}

An astronomical instrument and its calibration software are like partners in a marriage: they evolve over the years and must always stay in sync. Once the top-level science requirements are fixed, the marriage starts in the design phase. As detailed specs of instrument hardware change the specs of the software to calibrate the instrumental fingerprint need to change along. The marriage continues during the phases of commissioning, science verification and, regular operations. During the first two the difference between expected and real instrument behavior often requires significant updates to calibration procedures and hence software recipes. In regular operations the calibration scientist gains insight through trend analysis over years of observations and finetunes calibration software recipes to maximize scientific data quality.

\section{Challenge: calibrating MICADO at the ELT}

The Extremely Large Telescope (ELT, see https://www.eso.org/sci/facilities/eelt/) is ESO's 40 meter optical/near-infrared adaptive optics (AO) assisted telescope under construction on Cerro Armazones in Chile. This combination of size and AO opens up a part of parameter space of resolution and sensitivity for novel science.
It is an actively controlled system including a primary mirror of almost a 1000 alignable segments and two or more deformable mirrors. Its First Light imager and spectrograph  MICADO is a gravity invariant instrument, with 3 imaging modes, 5 movable wheels for filters and masks, an Atmospheric Dispersion Corrector, and 9 H4RG detectors \citep{2018SPIE10702E..1SD}.
The data flow of MICADO can reach $10^4$ raw exposures, or 6.7 terabyte, in 24hrs. The unprecedented combination of 'activeness' and size is challenging to calibrate. Yet the aim is to minimize time dedicated to calibration to maximize the novel science return. This demands a joint design phase and tight coupling between hardware specs and calibration software.

This leads to the following requirements on the process itself in the detailed design phase of instrument hardware, calibration plan and calibration software recipes:
\begin{itemize}
	\item {\bf Continuous iteration between hardware and calibration software team}: 
	As detailed specs of hardware change a (fast) iteration is needed to check if its instrumental fingerprint is calibratable, to detect flaws in the hardware plus software design and to trade-off between challenging manufacturing specs on hardware versus extensive calibration procedures using the data.  
	\item {\bf Software design digestible for hardware team}: the coupling between the designs of hardware and calibration software means the software design must also be understandable to the hardware team to obtain their valuable feedback on the bits that most intimately depend on the hardware. It should provide them both a high level overview of our pipelines and also allowing a dive into the instrument dependent details.
	\item {\bf Easy prototyping}:
	for the calibration software team the design should trivially translate into skeletons for the software prototypes and to connect it to the simulations. Prototyping should primarily be about developing and testing algorithms, while having a machine readable design framework that provides the prototype its specification of hardware specs, data specs, recipe inputs/outputs as automated and as easy maintainable as possible.
\end{itemize}

\section{Solution: a data-model-centric MICADO Pipeline Design}

Our solution for MICADO is to create a design specification framework that generates as automatically as possible not only the software prototype skeleton but also the human readable documentation. At the heart of the framework is a data model (see figure \ref{fig_workflow}):

\begin{itemize}
\item \textbf{Data Model}:
We follow a data-centric approach, so central to our workflow (for both design and processing) is a data model.
The data model describes the five major components of our design:
calibration requirements,
raw (calibration) observations,
instrument hardware specifications,
processed data products,
and software recipe specifications.
The hardware and software specification can effectively be seen as a property of the raw and processed data.

The data model is encoded in the IVOA standard VO-DML/XML to ensure compatibility with other software stacks (Virtual Observatory Data Modeling Language in XML, \citet{2018ivoa.spec.0910L}).
The VO-DML/XML is generated from VO-DSL (\url{https://github.com/pahjbo/vodsl}), an easier to write (and read) domain specific language.

\item \textbf{Prototypes}:
The data model is automatically converted to software skeletons, in our case Python classes.
Developers can fully focus on writing prototype algorithms, since all the boilerplate is generated automatically.
The Python code will be used as basis for our future consortium science platform \citep{2012ASPC..461..881B}.
The Python classes also form the glue to generate all the other output of the workflow.

\item \textbf{Data Reduction Library Specification and Calibration Plan}:
All figures and tables in the design documentation are automatically generated from the data model by using ditaa (\url{https://github.com/stathissideris/ditaa}), restructured text and Sphinx (\url{https://www.sphinx-doc.org}).
This ensures that everything in the design documents is both consistent with other parts of the documentation and with the prototypes.
More importantly it fosters efficient collaboration by making iterations seamless.

\item \textbf{Simulator Interface}:
The data model (the hardware specification in particular) is automatically converted into YAML objects to interface with the MICADO simulator \citep{2019ASPC..521..527L}.

\item \textbf{FITS Headers}:
Everything in the data model can be converted to ESO compliant FITS headers \citep{2011ASPC..442...33V}.
This ensures full provenance or data lineage for both simulated raw data as well as processed data.

\item \textbf{Database}:
The Python skeletons are based on the WISE technology by OmegaCEN \citep{2012ASPC..461..719V}.
The WISE technology includes an ORM that ensures everything can be stored persistently in a database.
The database will be used as simulation archive, and for the consortium science platform

\end{itemize}

\articlefigure[width=\linewidth]{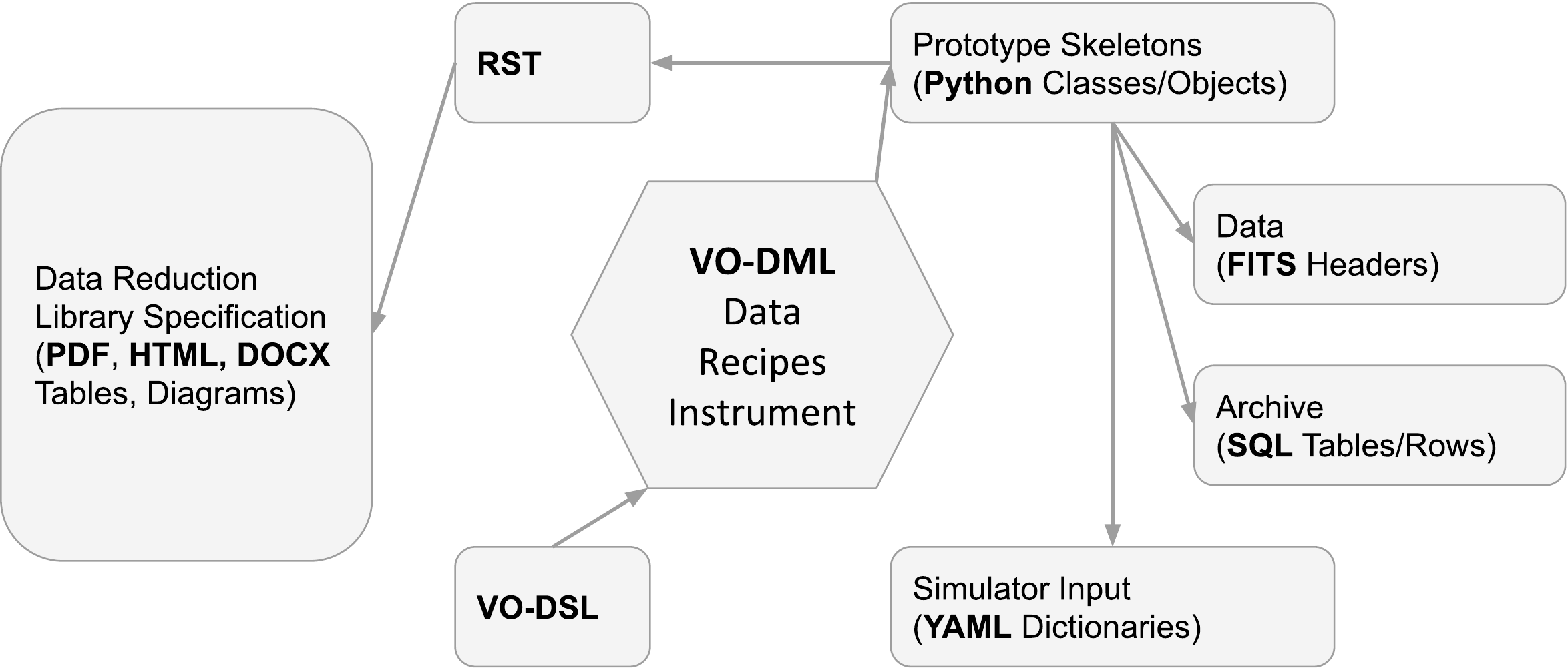}{fig_workflow}{
	Our design workflow.
	Boxes are representations of the model.
	Arrows are fully automatic conversions.
	Only the most important representations and conversions are shown; for example several conversions can also be inverted.
}

\section{Conclusions: an integrated hardware and software design process}

Our data-model-centric software design and calibration plan provides an agile and consistent method to deal with the evolution of instrument and calibration software.
For MICADO it integrates specification of the data simulator and pipeline prototype.
There are several components in the atmosphere, ELT, AO plus MICADO system that require complex calibrations.
At the same time the cost of the system puts pressure to minimize dedicated calibration time.
Our design philosophy ensures we can meet these conflicting demands by calibrating the instrument with maximum use of science data.

We gratefully acknowledge that our data-model-centric approach for MICADO is based on expertise and lessons-learned from earlier projects starting with the Odoco\footnote{https://www.eso.org/sci/facilities/paranal/instruments/omegacam/doc/DFS-CAL.pdf} system for OmegaCAM and Astro-WISE, followed by MuseWISE \citep{2015scop.confE..28V} and the Euclid Science Ground Segment.

Our approach to the design of the MICADO imaging pipeline shows that it is possible to have a machine readable design with off-the-shelf open source tools and standards.
The main benefits for us are extreme flexibility, implicit versioning, automatic consistency and minimal manual labor.
We encourage current and future software designers to improve on our approach to fit their needs.
Project managers are encouraged to promote open standards (e.g. rst/pdf, not Word).
As a community we can develop best practices for the design of future telescope software.

In the future we will use the prototypes as basis for MicadoWISE, our consortium science platform.
The final ESO-compliant algorithms will be a drop-in replacement for the prototype algorithms due to python-cpl bindings (\url{https://github.com/olebole/python-cpl}).
This will allow calibration scientists to gain insight by trend analysis over years of observations.
Furthermore the WISE technology generates processing workflows that optimize data reprocessing automatically.

\bibliography{O3-2}

% if we have space left, we might add a conference photograph here. Leave commented for now.
% \bookpartphoto[width=1.0\textwidth]{foobar.eps}{FooBar Photo (Photo: Any Photographer)}

\end{document}